# Google, AI Literacy, and the Learning Sciences: Multiple Modes of Research, Industry, and Practice Partnerships

Victor R. Lee (co-chair), Stanford University, vrlee@stanford.edu
Michael Madaio (co-chair), Google Research, madaiom@google.com
Ben Garside, Raspberry Pi Foundation, ben.garside@raspberrypi.org
Aimee Welch, Google DeepMind Impact Accelerator, aimeewelch@google.com
Kristen Pilner Blair, Stanford University, kpilner@stanford.edu
Ibrahim Oluwajoba Adisa, Stanford University, jobaa@stanford.edu
Alon Harris, Google Research, alonharris@google.com
Kevin Holst, Google Research, kevinholst@google.com
Liat Ben Rafael, Google Research, liatb@google.com
Ronit Levavi Morad, Google Research, ronitlm@google.com
Ben Travis, Phantom LLC, ben@phntms.com
Belle Moller, Phantom LLC, belle@phntms.com
Andrew Shields, Phantom LLC, andrew.shields@phntms.com
Zak Brown, Phantom LLC, zak@phntms.com
Lois Hinx, Phantom LLC, lois@phntms.com
Marisol Diaz, Massachusetts Institute of Technology, marisold@mit.edu
Evan Patton, Massachusetts Institute of Technology, ewpatton@mit.edu
Selim Tezel, Massachusetts Institute of Technology, stezel@mit.edu
Robert Parks, Massachusetts Institute of Technology, rparks@mit.edu
Hal Abelson, Massachusetts Institute of Technology, hal@mit.edu
Adam Blasioli, New York Jobs CEO Council, ablasioli@jobscouncil.org
Jeremy Roschelle (discussant), Digital Promise, jroschelle@digitalpromise.org

**Abstract:** Enabling AI literacy in the general population at scale is a complex challenge requiring multiple stakeholders and institutions collaborating together. Industry and technology companies are important actors with respect to AI, and as a field, we have the opportunity to consider how researchers and companies might be partners toward shared goals. In this symposium, we focus on a collection of partnership projects that all involve Google and all address AI literacy as a comparative set of examples. Through a combination of presentations, commentary, and moderated group discussion, the session, we will identify (1) at what points in the life cycle do research, practice, and industry partnerships clearly intersect; (2) what factors and histories shape the directional focus of the partnerships; and (3) where there may be future opportunities for new configurations of partnership that are jointly beneficial to all parties.

## Overview

In line with the theme of "Partnering with Purpose", we invite the field to consider what purposeful partnership looks like at our current time between industry and the learning sciences. While much work from the learning sciences community currently involves generative collaborations with organizational units such as schools, districts, informal learning spaces, civic groups, and other types of nonprofits, it is worth noting that there has been precedence for and interest in industry and learning sciences collaboration. For example, the Institute for the Learning Sciences at Northwestern University, founded in 1989 and considered an important catalyst for learning sciences in North America, was seeded by funding and collaboration with the business entity that was then called Andersen Consulting. This contributed to the design and development of goal-based scenario software environments (Schank et al, 1994) that integrated key discoveries about learning with job contexts like being a journalist or a sickle cell genetic counselor. Boeing, the aircraft manufacturing company, partnered with learning scientists at the University of Washington to develop and experimentally compare the effectiveness of new forms of employee training (O'Mahony et al., 2011). Now, with substantial investment, attention, and energy being directed to artificial intelligence in education, we believe a new conversation is needed. There are shared interests and concerns about achieving impact for learners and supporting future workforce preparation, where AI is expected to play a major role in our lives. At the same time, there are seemingly different orienting frameworks, incentives, and practices across the sectors of academic research and the tech sector. Partnerships between learning sciences research and industry are forming and maturing, with implications for practitioners. What can we learn from examining and contrasting examples of those?

This symposium brings together four projects that share three elements. One is that all the projects share a common industry partner: Google. While Google is so well known from its origins as a search engine that "google" is now a dictionary verb for online information search, the company is a highly diversified multinational corporation and subsidiary of Alphabet Inc. Service, with product teams that work on familiar consumer-facing tools and platforms like Google Suite, YouTube, Google Mail, FitBit, Chrome, Android, and Google Drive. Important for this symposium are units at Google that work on research and development, philanthropy, and education. As will be discussed in the symposium, there are many such units, and partnerships between non-Google researchers and educators may begin with one Google sub-unit but over time connect more with others. Moreover, funding for collaborations may come out of distinct units under different auspices tied to the unit's missions. In the symposium, we will critically reflect on and discuss the implications for learning scientists of partnerships with large corporations, particularly when those entities, in some of the projects, serve dual roles as both funder and research and development partner.

Another shared element across partnerships is that they grapple with ideas, questions, and decisions that are core to the ongoing intellectual dialogues of the learning sciences. For example, one project engages with questions of curriculum design and teacher learning in service of enacting curriculum, active concerns for the field (e.g., Edelson et al., 1999; Davis et al., 2017). Another is rooted in constructionist principles and how AI literacy can be fostered through environments oriented toward computational action (e.g., Kafai et al., 2024; Tissenbaum et al., 2019). One more project builds on scholarship around learning game design and digital scaffolds (e.g., Holbert & Wilensky, 2017). Additionally, one last focal project for the symposium focuses on city-scale learning, a growing area of interest in the field, as partnership efforts with community organizations are helping to shape learning opportunities for whole geographic regions. These are not mutually exclusive concerns across our contributors, especially as new directions for further collaboration among these projects or with additional partners are explored. We see the diversity of topic areas, learning theories, and learner communities as a strength of this symposium, as it allows us to make connections and discuss implications with relevance to a range of researchers and practitioners.

The third shared element is that the partnerships in this session all seek to address the development of AI literacy - a level of understanding about AI that does not presume learners will pursue future technical activities of AI software development, but instead that they will be able to effectively use or critically engage with AI thoughtfully and responsibly in their lives. Enabling AI literacy is a growing global concern and one that is being explored in the learning sciences (Kafai et al., 2024). Some efforts have been undertaken to define or represent it through the creation of various frameworks, formation of expert groups, or academic literature reviews (Long & Magerko, 2020; Miao & Shohira 2024; Mills et al., 2024; Ng et al., 2021). Still, to do work that can scale is a complex challenge requiring multiple stakeholders and institutions collaborating together. That includes collaborations that involve major tech companies that are creating new AI tools alongside learning scientists who use the tools of research and design to understand and improve human learning. This effort, of course, creates important opportunities to interrogate the extent to which we want to actively promote and scale technologies like AI and what boundaries need to be observed, particularly given the power dynamics involved in collaborations between learning scientists and tech companies (Nichols et al., 2025).

In this symposium, we will reflexively and critically examine and consider similarities and differences with respect to how the partnerships originated and evolved, integrated research and technical expertise, and generated impact in service of AI literacy. Such a reflection may gain new insights into the nature of partnerships that leverage the private sector and learning scientists. It would also expose lessons learned due to the complexities involved in developing AI literacy resources and experiences for different formats (e.g., curriculum, digital game, online interactive tool), audiences (e.g., middle school students, high school students, college students, and adults), and contexts (e.g., formal and informal learning, UK and US schools and the Global South). Through the session, we will identify (1) at what points in the life cycle do research, practice, and industry partnerships intersect; (2) what factors and histories shape the directional focus of the partnerships, including the myriad roles for learning scientists in partnerships with corporations; and (3) where there may be future opportunities for new configurations of partnership in the future that are jointly beneficial to all parties.

## AI literacy and the learning sciences

Given the increasing ubiquity of AI applications in public life, researchers and policymakers alike have called for supporting AI literacy in the broader public (Long & Magerko, 2020; Kafai et al., 2024; Miao & Shohira, 2024). In education, researchers have argued that improving AI literacy can lead to more effective and responsible use of AI to support learning (Kim et al., 2025; Humburg et al., 2025). The specific definitions of AI literacy vary, but they broadly include a set of competencies that enable people to critically reflect on how AI models and systems work, how they are designed, and thus, how they are (and should be) used - or not used

(Long & Magerko, 2020). Across various AI literacy frameworks, multiple perspectives on AI are promoted that differentially emphasize understanding the technical functionality of AI systems like training machine learning models or the use of sensors; human-centered, sociotechnical aspects of AI, such as the ways that diversity (or the lack thereof) in training data can shape the outputs of AI systems; the active role of humans in designing and developing AI models and AI-powered applications; how AI is should and could be used in different work functions and professional activities; and the positive and negative impacts of AI on society (Miao & Shohira, 2024; Lee & Long, to appear).

AI literacy has been a topic of interest to the ISLS community in recent years, with multiple symposia exploring different approaches to teaching and learning AI literacies in K-12 education (Kafai et al., 2024) and investigating how to incorporate equity in AI literacy programs (Walsh et al., 2022). Meanwhile, other learning sciences research has explored different aspects of AI literacy, including the impact of AI literacy instruction on students' self-efficacy (Tatar et al., 2023) and empowerment (Kramarczuk et al., 2025), as well as research on fostering teachers' AI literacy (Ocak et al., 2025) and how teachers integrate AI literacy into their lessons (Adisa et al., 2025), among others (e.g., Relmasira et al., 2024). This body of research continues to grow.

However, many current AI literacy educational experiences face systemic challenges. How and where AI literacy resides in the educational ecosystem remains unanswered, with crowded curricular requirements in schools, unequal access to physical and digital learning resources, and heterogeneity of the learner population - ranging from students, educators, and adult learners - each with their own distinct learning needs.

## Symposium Format

The symposium is structured to be a joint sense-making session for participants and attendees rather than a series of collected presentations. It will involve short lightning talks from invited speakers about each of the partnerships described below, all speaking to the symposium theme about partnerships that enable AI literacy. These will be relatively short so that there can be commentary from the invited discussant, Jeremy Roschelle, Co-Executive Director and co-lead of Learning Sciences Research at Digital Promise. Roschelle will offer his perspective on the nature of these partnerships and what we might expect in the future for private sector and researcher collaborations. Following that, the bulk of the time is intended to be a Q&A discussion from the audience that will be moderated by Roschelle, who will also make thematic connections as they appear in the discussion across each of the presentations, with implications for industry partnerships beyond Google.

## Raspberry Pi Foundation: Training teachers and reaching global learners with Experience AI

Raspberry Pi Foundation (RPF): Ben Garside
Google DeepMind Impact Accelerator (GDI): Aimee Welch

## Overview

Raspberry Pi Foundation is a non-profit organization in the United Kingdom that promotes computer science education worldwide. Its reputation and name as a leader in this space was established from the launch of the Raspberry Pi single-board controller that gained popularity as part of the Maker movement and for Internet of Things applications. Google's DeepMind is a leading AI lab that emphasizes work that supports scientific breakthroughs. Some of these include AlphaGo, the first computer program that defeated a Go world champion, transformer architectures that are used in modern language models, and AlphaFold, which predicts the structures and interactions of proteins, DNA, RNA, ligands and more with high accuracy. DeepMind is also behind Gemini, the family of AI models used across Google text, code, image, audio, and video services.

*Experience AI* is a free educational program, released in 2023 and developed in a partnership between the Raspberry Pi Foundation and Google DeepMind. It offers instructional resources on artificial intelligence (AI) and machine learning for educators and young people aged 11 to 14 related to the workings of AI, such as lessons on bias, decision trees, and large language models. The *Experience AI* theory of change is pursued through three main activities: providing high-quality, research-led learning content; localizing and translating content for market relevance; and partnering with expert implementers to train educators.

A significant global challenge in introducing AI literacy to classrooms is teacher preparedness. Research (Rajapakse et al., 2024) indicates that the rapid pace of change in AI leaves educators feeling unready to teach the skills necessary for future generations. The program aims to bring AI literacy education to young people globally by upskilling educators, providing them with the resources and self-efficacy to support learners in understanding AI's potential role in their lives. Recognizing that there are many specific circumstances for different parts of the world that look to Raspberry Pi Foundation for educational support, they have additional collaborations and partners (27 partners in 25 countries) that work to train educators and build their confidence in teaching about AI to their students. The team and their partners focused on cultural relevance for the many locations and communities involved - this involves not only translating resources for educators and learners worldwide but also identifying points and examples of local relevance to help make connections to AI more visible. *Experience AI* is presently estimated to have reached 2 million learners, primarily aged 11-14, although a number of *Experience AI*'s partners are working with young people up to the age of 18. They prioritize working with partners in low and middle-income countries whose mission includes reaching educators and young people in underserved communities. Foci include foundational AI concepts, understanding of AI's impact on the world, awareness of AI-related careers, and comprehension of societal and ethical issues. Training is delivered to educators via both synchronous sessions and a free asynchronous online course.

**Figure 1**
Raspberry Pi Foundation partners with educators and young people across the globe.

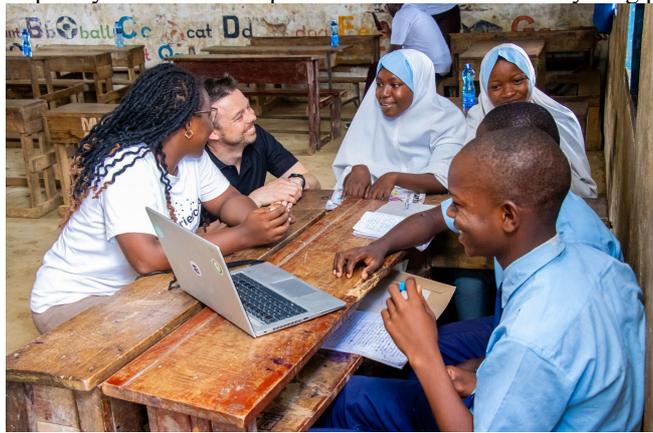

## Partnerships

*Experience AI* (experience-ai.org) was built through a collaboration between industry experts and researchers at the Raspberry Pi Foundation and the University of Cambridge, bringing together cutting-edge practice and academic rigor to ensure that content and pedagogy are evidence-led (Waite and Garside, 2023; Waite et al., 2024). Google DeepMind provided initial funding for *Experience AI* lesson resources and crucial guidance from industry experts for the team at Raspberry Pi Foundation, who led content and learning resource development. This partnership thrived as Google DeepMind offered real-world AI contexts, case studies, and examples, enabling young people to understand the tangible impact of AI as a problem-solving tool.

With the goal of democratizing access to AI literacy and to make visible that AI education is relevant to all, *Experience AI* lessons intentionally featured aspirational role models from a wide range of backgrounds involved in developing AI systems at Google DeepMind. Furthermore, Google.org, the philanthropic arm of Google, has facilitated the global expansion of Experience AI by providing grant funding for Raspberry Pi Foundation to scale the curriculum by taking a coordinating role and supporting 27 partners across 25 countries since 2024 (with plans to add more in 2026). These partners—ranging from non-profits and charitable foundations to socially-driven private companies—receive training in AI literacy concepts as well as access to the *Experience AI* materials. Partners work with the *Experience AI* team to make lesson adaptations so the contexts and examples used in the materials are relevant and relatable to young people in their region. In this expansion, it was essential to develop collaborations with in-country partners who have a deep understanding of their educational landscapes.

## Insights

One key insight has been that while the *Experience AI* "Foundations" unit was originally designed for a context (UK) where a national curriculum for computer science creates a space for teachers to freely introduce AI literacy—this is not the case in most of the countries who use *Experience AI*. Many do not have a dedicated computer science curriculum, which means teachers often struggle to see where AI fits. This surfaces the need for the team to provide stronger guidance and signposting to show how E*xperience AI* lessons connect with a range of subject areas. In addition, AI literacy resources are now being created that are tailored to specific subjects, so that they can be more easily integrated into existing curricula and classroom practice, similar to the "across the disciplines" approach advocated by Jiang et al (2022).

One example of this has been a supplemental stand-alone lesson on "AI and Ecosystems" designed to be taught in a science class. A key part of the process was to partner with subject matter experts from the UK's Royal Society of Biology Association's Biology Education Research Group. From this, design insights have been made regarding the importance of: 1) disciplinary experts; 2) precise vocabulary; and 3) active consideration of existing disciplinary curriculum requirements.

Another important insight from work in low- and middle-income countries is that access to technology, the internet, and even electricity is highly varied. To ensure that students in these contexts are not further disadvantaged, the team has made provision for "unplugged" activities. These enable learners to meet the same learning objectives without relying on technology, supporting AI literacy education in a way that can be more inclusive and accessible regardless of local circumstances.

## *AI Quests*: An educational game for AI literacy

Google Research: Ronit Levavi Morad, Alon Harris, Liat Ben Rafael, Michael Madaio, Kevin Holst
Stanford University and Stanford Accelerator for Learning: Victor R. Lee, Kristen Pilner Blair, Joba Adisa
Phantom LLC: Ben Travis, Belle Moller, Andrew Shields, Zak Brown, Lois Hinx

### Overview

In 2024, Google Research conducted market research regarding in-class and on-line AI literacy programs and curricula, and identified a critical gap in the market: very few programs addressed applied AI and its real-world impact. Members of that team then asked how they could leverage the impact-driven AI research across climate, health, and science to address this gap in an engaging way for teens—the answer: *AI Quests* (Figure 2). *AI Quests* is a free, in-class, online and game-based digital learning experience designed to enhance AI literacy for middle school students (ages 11-14). Developed by Google Research in partnership with the Stanford Accelerator for Learning & Phantom Studios, it provides an accessible entry point into applied AI. After launching in September 2025, over 200,000 people completed AI quests in the first 6 months of release.

**Figure 2**
Introductory screen for AI Quests describing the digital context.

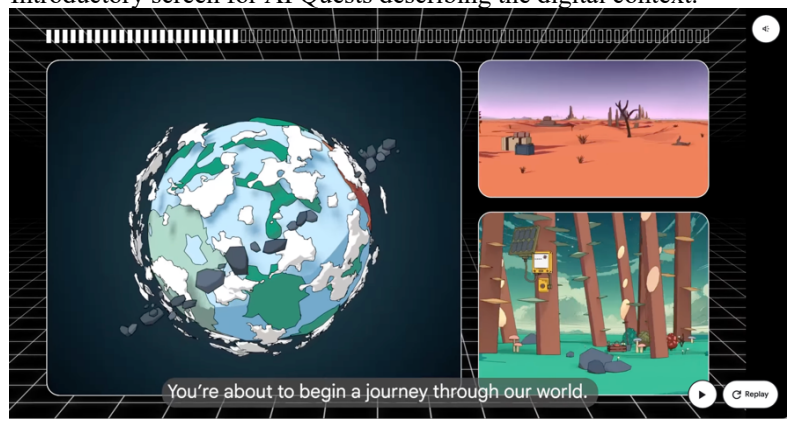

In *AI Quests*, students explore the fundamentals of AI and its applications through a series of immersive adventures in a fantasy world where students face compelling problems inspired by genuine research projects at Google Research, tackling issues like climate change and public health. The goal of *AI Quests* is to empower the youth to understand how humans can design AI to solve some of humanity's biggest challenges.

In the first quest, inspired by published "Flood Forecasting" research (e.g., Nevo et al., 2022; Nearing et al., 2023), students learn how AI can help predict floods by collecting relevant data sources, like rainfall and river flow, evaluating the quality of the data, and training and testing their model. Through various choice-based game mechanics, students learn how factors such as human judgment and data quality shape AI's outputs and impacts—critical competencies of AI literacy (Long & Magerko, 2020). Another quest launched in December 2025, which draws inspiration from an AI model developed by Google Research to detect a condition called diabetic retinopathy, which can lead to blindness (Bora et al., 2021; Varadarajan et al., 2020). A third quest will be released in 2026, which takes inspiration from "Connectomics" research to map and understand the human brain (Januszewski et al., 2018; Dorkenwald et al., 2023; Shapson-Coe et al., 2024). The benefit of using Google Research projects is the accessibility of experts who were part of the actual projects who can consult with the team on "what really happened" and help evaluate the ways in which the process of developing and deploying new AI in these contexts are represented.

At the end of each quest, students receive a recorded video message from the real researchers behind the work about how they are using AI responsibly to address challenges faced by millions of people worldwide. This provides a different image of AI that goes beyond chatbot technology and represents what happens in AI research that works for social impact. Each quest also comes with a complete lesson plan and teacher guide, with pre- and post-game activities.

Designed as a flexible, modular add-on to existing AI literacy curricula, *AI Quests* serves as a tool for educators and AI literacy programs seeking to introduce students to the practical application of AI and the reflective and sociotechnical thinking required in its development.

## Partnerships

For the realization of our vision for *AI Quests* we created a partnership "triad" between Google Research, Stanford Accelerator for Learning (SAL), and Phantom Studios. SAL (USA) is a university-housed entity that includes faculty, staff, and students in partnership and practice-focused collaborations as part of university effort to further tighten the links between research - including specifically from the learning sciences - and educational practice. Phantom Studios is a UK-based creative agency that develops digital experiences for clients and firms.

Google Research funded and managed the project as a whole, selected the research projects, onboarded the respective Google research teams, and owned the product management and scaling strategy. SAL drove the pedagogy of both the online game experience as well as in-class teacher facilitation of that experience through the design of curriculum materials. Phantom drove the creative translation of the research subject matter into engaging and age-appropriate digital experiences.

While there had been interest in helping students to see the nature of actual AI research in the context of social impact projects, there were clearly many possible new ideas for students to encounter. Through the partnership, a process for articulating learning goals and the overarching design process was established based on Backwards Design (Wiggins & McTighe, 2005). As a new framework for some partners, this approach introduced from learning scientists helped to articulate the nature of the human-centered focus - suggesting some backgrounding of technical terminology of AI that was deemed important to know but not part of the enduring understanding targeted.

The partnership also allowed for further investigation of some of the game mechanics and story sequence design that had already been proposed and drafted. Learning sciences research and key principles were then introduced to frame discussion of game mechanics, ultimately leading to specific strategies such as pedagogical agents - including teachable agents specifically (Blair et al., 2007) - and also encouraging interactions between player and characters that were inspired by the self-explanation effect (Chi et al., 1994). Also key was the decision to create adaptable and educative lesson materials for teachers who would also be learning the content and learning pedagogy. Through these conversations, the partners began to establish a general framework for quests that include specific recurring types of pedagogical agents, specific features and representations for depicting data used in training of AI and the machine learning process, reflection and interaction types, and internal game structures that are internally called "learning tickets" as they function as being like an exit ticket but are also intentionally moments that push for player reflection. Phantom Media had engaged in continuous player testing as the learning game experience went from storyboard to a progressively refined interactive digital experience.

## Insights

Consistent with emerging literature on partnership research, there was multidirectional learning throughout the entire team (Goldman et al., 2022). Co-design took place through weekly meetings of the "triad" of Google Research, SAL, and Phantom Media. Each partner contributed distinct and essential expertise needed to bridge the AI literacy gap. The integration of these roles - industry expertise, learning sciences rigor, and creative production - was necessary to design and iteratively improve the experience as ideation moved into prototype and ultimately into the final product.

The partnership introduced constraints that ultimately fostered creativity and productive compromise. One constraint has been the technical subject matter, drawing from real Google Research projects. Devising approximations of processes and practices that are accurate but intelligible has been an iterative process and also required continual return to backwards design principles. This is also shaped by knowledge about the target learners - middle school students aged 11–14 - learning in a classroom context. There were some prior intuitions and play patterns (such as simply clicking through rather than engaging with the content) that needed to be observed and addressed strategically using expertise brought from all parties. Importantly, production constraints factored in with an accountability to release products that reflected the professional expertise that had been put into place, but had legitimate timeline and budgetary constraints that required compromises on what could be pursued. For example, some important learning effects can be realized in a one-on-one situation in a lab but would not be scalable without a trained researcher present - therefore, approximations had to be designed. In many respects, this reflects some of the inherent challenges associated with design-based research in which real contexts shape what knowledge is practical and useful for the design task.

The workflows to which each party was accustomed had to be renegotiated. For example, production teams value clear and complete specifications so they can organize their work internally. However, a collaboration like this in which many prototypes were subject to revision on the basis of different expertise (how the real-world project in which a quest was based is best represented, or what hidden caveats or considerations had to be integrated into a designed learning feature for it to be effective) necessitated more production iterations than expected.

## *MIT App Inventor*: Student-directed AI literacy learning through tool-building
MIT: Marisol Diaz, Evan Patton, Selim Tezel, Robert Parks, Hal Abelson

### Overview
One core perspective for the design of digital learning experiences in the learning sciences has been Constructionism (Papert, 1980). Constructionism posits that learning takes place felicitously with the creation of a personally meaningful public artifact. Therefore, new expressive and digital media that allow for the generation of artifacts are especially valued so long as they provide opportunities for students to tinker, explore, create, and share their work and experiences.

*MIT App Inventor* is a platform created with constructionist principles in mind (Pang et al., 2023). It leverages block-based coding environments, enabling students to build their own apps that run on mobile devices. Computational empowerment and computational action (Kafai et al., 2024) are primary platform goals; the platform developers and project leaders aspire to empower students to use computational tools and media in ways that help address local concerns and bring about change for community betterment. As such, many users of *MIT App Inventor* are located around the world, with a large and growing user base in the Global South.

While *MIT App Inventor* originated from broader aims to support computational literacy broadly, AI literacy has been embraced as highly relevant and aligned with the history and aims of the platform, which has longstanding connections to MIT's CSAIL (Computer Science and Artificial Intelligence Laboratory). Google has been a key resource for the platform, given the AI tools it has created. Now, students can create apps embedded with Gemini chatbots, Google *Teachable Machine* image classification (von Wangenheim et al., 2024), and other embedded mobile AI tools.

Last year, students built over 300,000 novel apps embedded with AI functions in *MIT App Inventor*. The target population had been educators and students in grades 5-12, although some users are as young as 8 years old, and some users are non-educator adults. The overall makeup of users is 84.9% students, 7.3% teachers, and 7.8% adults.

Globally, teachers have used the free, open-source platform as a resource for student-directed learning in AI and coding, with 48% of *MIT App Inventor* users in classrooms in the Global South. The *MIT App Inventor* site is automatically translated for local use into 19 languages using Google web tools. For example, over 25.6% of our users are seeing the coding platform in Spanish. Students in Mexico are the 3rd largest user base at (8.4% of student users vs. 12. 9% US). Brazilian students (3.6%) are reading the site in Portuguese.

### Partnerships
Historically, Google has been a significant collaboration partner for *MIT App Inventor*, as the tool was incubated at Google during the project PI's (Hal Abelson) academic sabbatical in 2010 with Google. Abelson then launched *MIT App Inventor* in collaboration with Google soon after. Some of the partnership work has involved supporting the integrations with Google products, such as the Android operating system for mobile devices, and using Google tools like Gemini and *Teachable Machine*.

A specific new initiative that further extends the relationship between *MIT App Inventor* involves working with more Latin American and Brazilian students (Figure 3). Google for Education LATAM provided support to initiate relationships with Latin American ministers of education. These ministers and Google for Education representatives then participated in the 2025 MIT AI + Education Summit. Through that summit, LATAM ministers were able to see projects created by young people from around the world and learn about programs and offerings affiliated with the *MIT App Inventor* team. The ministers have stated that they gained first-hand knowledge of AI literacy and new ideas for promoting student-directed app-making in schools. With the help of the Google for Education team, they have started to plan another initiative for the Education + AI Summit in 2026. Convenings such as these represent an important collaboration and direction for partnership, particularly as they gain the attention of decision-makers and leaders. Other units and representatives in Google who focus on Latin America, such as team members from Google Workspace and Gemini for Education, also participated in the summit and conversations. In this regard, the nature of partnerships can span across teams in a single organization if they share a common focus (*e.g.*, engaging with Latin America).

**Figure 3**
Teachers from Puerto Rico training about teaching AI literacy through mobile app-building to educators from Latin America.

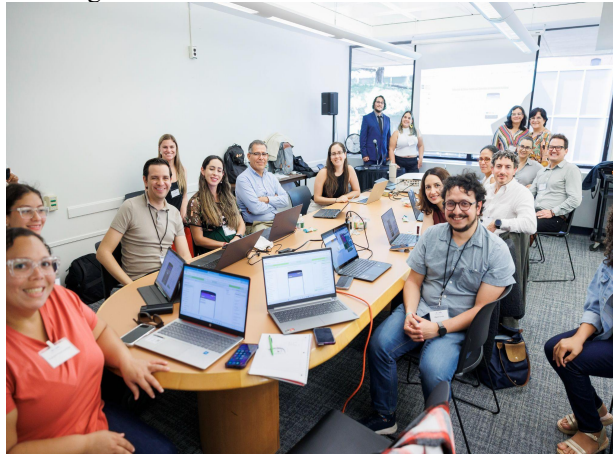

It is important to acknowledge, however, that while Google has been an important partner, many partnerships and collaborations have been critical, such as with the App Inventor Foundation, Brazilian Creative Learning Network, CienciaPR, and the Scientific Caribbean Foundation. The larger network plays an important role in engaging with different audiences and providing resources for sustainability.

## Insights

The story behind the collaboration may help researchers see how pathways for such partnerships occur in the wild. The team started with the notion to provide workshops in Spanish and Portuguese at theirl MIT AI + Education summit. But who would come? Summit organizers, who included many *MIT App Inventor* personnel, reached out to Google for Education LATAM to help create a network to assist in this effort.

Shared interest from Google for Education LATAM led to monthly check-ins to provide feedback as the MIT summit program came together. Personal experience from having worked in Latin America on Google's behalf was shared, which informed how to enhance the international collaboration and provide a more enriching experience for the Latin-American ministers. For instance, some workshop program segments were restructured and extended in order to account for different backgrounds and likely questions from the LATAM education ministers.

Google provided both financial support and support in terms of international expertise to coordinate virtual educator PD sessions before the summit, as well as in-person visits to the summit from groups the App Inventor team has always wanted to reach, including the Brazilian Creative Learning network (BCLN), Puerto Rican educators in CienciaPR, and the Scientific Caribbean Foundation. For instance, in June 2025, the App Inventor team conducted two weeks of AI literacy PD sessions on the topic of AI for Impact. This is progressing to further collaborations. *The MIT App Inventor* team has since begun collaborating with BCLN to develop an App Inventor Day in Brazil in order to broaden outreach and participation.

## **New York Jobs CEO Council: A public sector partnership to enable AI literacy**
Jobs Council: Adam Blasioli
Google Research: Michael Madaio

## Overview

Learner engagement and opportunity coordination within multiple groups within municipalities is a growing area of interest for learning sciences research (e.g., Ching et al., 2015; Pinkard, 2019; Taylor & Hall, 2013). Ultimately, the challenge these researchers address is how to leverage the resources residing across sites in a city in a way that promotes coherence and increases learning opportunities for learners. In this session, we share a collaboration that involves Google Research and the New York Jobs CEO Council. The New York Jobs CEO Council (Jobs Council) is an initiative to coordinate and unify New York City employment needs by encouraging new hiring practices (e.g., competency emphasis rather than credential emphasis) and to establish new learning and training experiences for low-income job candidates to ensure they have current industry-relevant skills. Membership includes leaders and representatives across area companies who are major employers in the city and committed to democratizing access to employment opportunities that drive economic mobility. AI literacy in the NYC area is a recent area of concentration for the Jobs Council, and takes the form of a five-part, in-person program where local college students learn about the fundamentals of generative AI. The course material has been developed in partnership with Google Research and other Jobs Council member companies and is intended to both demystify the technology and define AI literacy as a critical workforce skill.

The primary objective of this partnership around AI literacy is to ensure City University of New York (CUNY) students have access to industry-backed training that provides a foundational understanding of GenAI and prepares job seekers for a rapidly evolving workplace (Figure 4). The target audience are CUNY students, particularly those outside of Computer Science majors.

**Figure 4.**
An in-person class hosted by the Jobs Council, covering the use of AI agents and Gen AI applications in the workplace

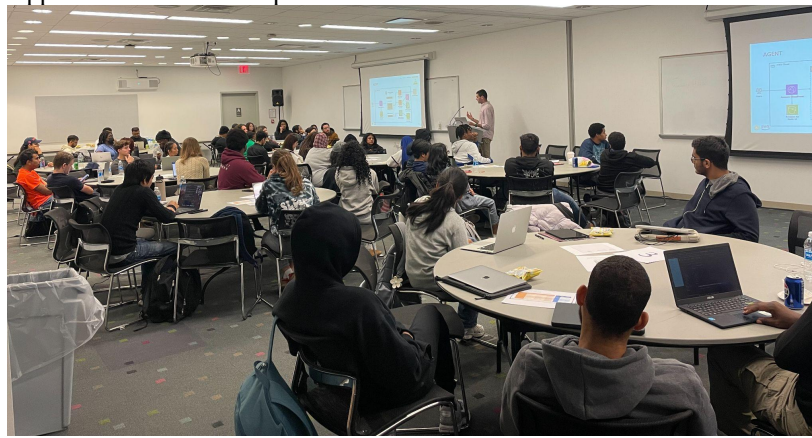

Many of the students who participate are (1) low-income, (2) first-generation college students, and (3) seeking guidance on how best to skill themselves around GenAI. A particular challenge for this population is that messaging from their school administrators and faculty is inconsistent around the usage of GenAI and has, at times, led to an erosion of trust between both parties. Not all students are gaining exposure or guidance on how to use AI for their academic experience nor how to prepare for future employment as AI changes the nature of work. The collaborators on this project work to develop students' capacity to understand and use the technology and to reinforce its application as a workforce skill necessary for future success.

Since the launch of the pilot in 2024, participants have reported an increase in their knowledge of GenAI, their confidence in using emerging tools, and their ability to explain the technology to others. More importantly, the series has provided a chance to address students' fears around the technology, including its perceived biases and potential to eliminate jobs. Additionally, the program's early successes have built momentum on campuses to explore opportunities to scale this work - including by embedding content into university curricula - and efficiently reach more students annually. Google's partnership in this effort has helped to embed practical industry expertise into the coursework, and the availability of their existing AI Essentials programming is offering a roadmap to expand and strengthen the delivery of this content to more students. Thus, earlier educational resources produced by other groups at Google, such as their public professional training subunit, "Grow with Google", serve as a resource for the curricular planning.

## Partnerships

The primary differentiator for this partnership in comparison to the other projects shared in this symposium is the heavy involvement of multiple private companies who are members of the Jobs Council.  The private companies, including Google, EY, JPMorganChase, Amazon, and others, have supported the development of content for the series and made employees available to lead training sessions, speak on panels, and meet directly with students through the coursework. The curricula were developed by them (e.g., Google AI Essentials, EY Skills Passport), their employees have led the sessions (e.g., EY, PwC, JPMorgan Chase, AIG, ConEd), and the collaborators at Google Research and the Jobs Council have adapted the offering in real-time to reflect students' needs and employer insights.

In addition to private companies, this project involves public sector partners and higher education institutions. One public sector partner has been the Tech Talent Pipeline, an entity supported by the NYC Mayor's Office, that has supported the promotion of this initiative and alignment to broader city-wide AI literacy efforts. Finally, the primary higher education partner has been CUNY which has both recruited students for this program, provided space to host the training sessions, and contributed to the identification of near-term needs and long-term scalable solutions. Each partner brings networks and specialized expertise to help aspects of the initiative align with practices and activities within other organizations.

## Insights

This project and its partnership offer a perspective on how higher education, government, and the private sector must work to coordinate goals and draw on complementary resources and expertise to create new pathways for learning at the city scale. This initiative sheds light on how public sector agencies (e.g., the Mayor's Office of Workforce Development, and the Tech Talent Pipeline) can leverage their comprehensive view of AI-related employment trends and needs for particular skills around AI literacy at the city-wide scale, to act as a hub to convene multiple private sector actors who may not otherwise have the capacity to act at a city scale.

Meanwhile, this partnership also provides insight into how private sector employees can leverage their expertise about AI in the tech sector to inform an AI literacy curriculum and training, including how AI is currently being used in industry software development, knowledge of the AI competencies they look for when hiring, and the soft skills, or sociotechnical knowledge and skills about using AI at work. However, this effort raises questions about how best to support industry tech workers who are not trained as teachers in being equipped to effectively teach or facilitate training sessions for AI literacy. Future versions of this AI literacy initiative are being planned, drawing on lessons learned from the first year of pilots, including a train-the-trainer model, recruiting CUNY alumni employed by partner companies, and plans to scale the program's reach.

## Significance of the symposium

As the learning sciences examines and expands practices of partnership in service of advancing research and improving practice, this collection of projects advancing AI literacy with a common industry partner, but differing strategies for learner engagement, provides a unique opportunity to draw comparisons and contrasts of relevance to the field and for future learning sciences work that connects to the private sector. For instance, while there is one prominent company - Google - that is involved, the ensemble of contributors shows how there are many different inroads for partnership made with many distinct sub-units housed in a single company. The global reach enabled by the company partnership allows for research that can have impacts across national boundaries, although that cross-national reach introduces new considerations and the need for even more collaborators. There are also important considerations of multiple goals and motivators, some aligned and some divergent, for designing and modifying learning experiences regarding AI literacy, worthy of consideration. Scale and networks of additional partners appear in this set of examples, and several of the contributions illustrate how the partnerships often require more than two partner organizations be involved. Ultimately, this collection of examples raises new questions for the field for what new knowledge is important to develop in order to enact change and to improve insight on how educational innovation is launched, adapted, and shared through private sector partnership, along with important considerations and tensions for future partnerships between learning scientists and other tech companies. In addition, we anticipate the attendees and discussion that takes place during the symposium amongst speakers and among attendees will uncover other notable qualities and features that can help all interested parties in developing partnerships that can simultaneously build knowledge and achieve impact.

## Acknowledgements
We want to thank all of our partners, as well as Yossi Matias for his strategic vision and support for AI literacy.

# References

Adisa, I. O., Mah, C., & Lee, V. R. (2025). High School Teachers' Approaches for Integrating AI Literacy into Planned Instruction: Uniformity or Heterogeneity?. In *Proceedings of the 19th International Conference of the Learning Sciences-ICLS 2025*, pp. 1545-1549. International Society of the Learning Sciences.

Blair, K., Schwartz, D. L., Biswas, G., & Leelawong, K. (2007). Pedagogical agents for learning by teaching: Teachable agents. *Educational technology*, 56-61.

Bora, A., Balasubramanian, S., Babenko, B., Virmani, S., Venugopalan, S., Mitani, A., ... & Bavishi, P. (2021). Predicting the risk of developing diabetic retinopathy using deep learning. *The Lancet Digital Health, 3*(1), e10-e19.

Chi, M. T., De Leeuw, N., Chiu, M. H., & LaVancher, C. (1994). Eliciting self-explanations improves understanding. *Cognitive science, 18*(3), 439-477.

Ching, D., Santo, R., Hoadley, C., & Peppler, K. (2015). *On-Ramps, Lane Changes, Detours and Destinations: Building Connected Learning Pathways in Hive NYC through Brokering Future Learning Opportunities*. Hive Research Lab.

Davis, E. A., Palincsar, A. S., Smith, P. S., Arias, A. M., & Kademian, S. M. (2017). Educative Curriculum Materials: Uptake, Impact, and Implications for Research and Design. *Educational Researcher, 46*(6), 293-304. doi:10.3102/0013189X17727502

Dorkenwald, S., Li, P. H., Januszewski, M., Berger, D. R., Maitin-Shepard, J., Bodor, A. L., ... & Jain, V. (2023). Multi-layered maps of neuropil with segmentation-guided contrastive learning. *Nature Methods, 20*(12), 2011-2020.

Edelson, D. C., Gordin, D. N., & Pea, R. D. (1999). Addressing the challenges of inquiry-based learning through technology and curriculum design *Journal of the learning sciences, 8*(3/4), 391-450.

Goldman, S. R., Hmelo-Silver, C. E., & Kyza, E. A. (2022). Collaborative design as a context for teacher and researcher learning: Introduction to the special issue. *Cognition and Instruction, 40*(1), 1-6.

Holbert, N., & Wilensky, U. (2019). Designing Educational Video Games to Be Objects-to-Think-With. *Journal of the learning sciences, 28*(1), 32-72. doi:10.1080/10508406.2018.1487302

Humburg, M., Han, A., Zheng, J., Rosé, C. P., Chao, J., Melo, N. A., ... & McBride, C. (2025). Humanizing AI for Education: Conversations with the JLS 2026 Special Issue Contributors. In *Proceedings of the 19th International Conference of the Learning Sciences-ICLS 2025*, pp. 2269-2277. International Society of the Learning Sciences.

Januszewski, M., Kornfeld, J., Li, P. H., Pope, A., Blakely, T., Lindsey, L., ... & Jain, V. (2018). High-precision automated reconstruction of neurons with flood-filling networks. *Nature methods, 15*(8), 605-610.

Jiang, S., Lee, V. R., & Rosenberg, J. M. (2022). Data science education across the disciplines: Underexamined opportunities for K-12 innovation. *British Journal of Educational Technology, 53*(5), 1073-1079.

Kafai, Y. B., Proctor, C., Cai, S., Castro, F., Delaney, V., DesPortes, K., ... & Rosé, C. (2024). What Does it Mean to be Literate in the Time of AI? Different Perspectives on Learning and Teaching AI Literacies in K-12 Education. In *Proceedings of the 2024 International Conference of the Learning Sciences*, pp. 1856-1862, International Society of the Learning Sciences.

Kim, J., Detrick, R., Lee, S. S., & Li, N. (2025). Impact of AI Literacy on Student-AI Interactions on Academic Writing. In *Proceedings of the 19th International Conference of the Learning Sciences-ICLS 2025*, pp. 1265-1269. International Society of the Learning Sciences.

Kramarczuk, K., Morey, J., Szabo, N., & Kedia, A. (2025). Empowering Undergraduate Computer Science Majors Through AI Literacy Research. In *Proceedings of the 19th International Conference of the Learning Sciences-ICLS 2025*, pp. 440-448. International Society of the Learning Sciences.

Lee, V. R., & Long, D. (to appear). AI literacy: Perspectives underlying an essential new digital literacy In J. Castek, J. Coiro, E. Forzani, C. Kiili, M. S. Hagerman, & J. R. Sparks (Eds.), *The International Handbook Of Research In Digital Literacies* New York, NY: Routledge.

Long, D., & Magerko, B. (2020, April). What is AI literacy? Competencies and design considerations. In *Proceedings of the 2020 CHI conference on human factors in computing systems* (pp. 1-16).

Miao, F., & Shiohira, K. (2024). *AI competency framework for students*. UNESCO Publishing.

Mills, K., Ruiz, P., Lee, K. W., Coenraad, M., Fusco, J., Roschelle, J., & Weisgrau, J. (2024). AI Literacy: A Framework to Understand, Evaluate, and Use Emerging Technology. *Digital Promise*.

Nearing, G., Cohen, D., Dube, V., Gauch, M., Gilon, O., Harrigan, S., ... & Matias, Y. (2023). AI increases global access to reliable flood forecasts. *arXiv preprint arXiv*:2307.16104.

Nevo, S., Morin, E., Gerzi Rosenthal, A., Metzger, A., Barshai, C., Weitzner, D., ... & Matias, Y. (2022). Flood forecasting with machine learning models in an operational framework. *Hydrology and Earth System Sciences, 26*(15), 4013-4032.

Ng, D. T. K., Leung, J. K. L., Chu, S. K. W., & Qiao, M. S. (2021). Conceptualizing AI literacy: An exploratory review. *Computers and Education: Artificial Intelligence, 2*, 100041.

Nichols, T. P., Logan, C., & Garcia, A. (2025). Generative AI and the (Re)turn to Luddism. *Learning, Media and Technology, 50*(3), 379-392. doi:10.1080/17439884.2025.2452199

O'Mahony, T. K., Vye, N. J., Bransford, J. D., Sanders, E. A., Stevens, R., Stephens, R. D., . . . Soleiman, M. K. (2011). A Comparison of Lecture-Based and Challenge-Based Learning in a Workplace Setting: Course Designs, Patterns of Interactivity, and Learning Outcomes. *Journal of the learning sciences, 21*(1), 182-206. doi:10.1080/10508406.2011.611775

Ocak, C., Han, S. C., & Caskurlu, S. (2025). Comparing Human and Machine Responses to Foster In-service Teachers' AI Literacies. In *Proceedings of the 19th International Conference of the Learning Sciences-ICLS 2025*, pp. 1829-1833. International Society of the Learning Sciences.

Pang, H.N., Parks, R., Breazeal, C., & Abelson, H. (2023). The effect of computational action on students' computational identity and self-efficacy, In *Proceedings of EDULEARN23*, pp. 8404-8412.

Papert, S. (1980). *Mindstorms: Children, Computers, and Powerful Ideas*. Basic Books.

Pinkard, N. (2019). Freedom of Movement: Defining, Researching, and Designing the Components of a Healthy Learning Ecosystem. *Human Development, 62*(1-2), 40-65. doi:10.1159/000496075

Rajapakse, C., Ariyarathna, W., & Selvakan, S. (2024). A self-efficacy theory-based study on the teachers' readiness to teach artificial intelligence in public schools in Sri Lanka. *ACM Transactions on Computing Education, 24*(4), 1-25.

Relmasira, S. C., Donaldson, J. P., & Lai, Y. C. (2024). Toward a Theory-Grounded Associative Model of AI Literacy. In *Proceedings of the 18th International Conference of the Learning Sciences-ICLS 2024*, pp. 2057-2058. International Society of the Learning Sciences.

Schank, R. C., Fano, A., Bell, B., & Jona, M. (1994). The design of goal-based scenarios. *Journal of the learning sciences, 3*(4), 305-345.

Shapson-Coe, A., Januszewski, M., Berger, D. R., Pope, A., Wu, Y., Blakely, T., ... & Lichtman, J. W. (2024). A petavoxel fragment of human cerebral cortex reconstructed at nanoscale resolution. *Science, 384*(6696), eadk4858.

Tatar, C., McClure, J., Bickel, F., Ellis, R., Wiedemann, K., Chao, J., ... & Rosé, C. P. (2023). Examining high school students' self-efficacy in machine learning practices. In *Proceedings of the 17th International Conference of the Learning Sciences-ICLS 2023*, pp. 1434-1437. International Society of the Learning Sciences.

Taylor, K., Headrick, & Hall, R. (2013). Counter-Mapping the Neighborhood on Bicycles: Mobilizing Youth to Reimagine the City. *Technology, Knowledge and Learning, 18*(1-2), 56-93.

Tissenbaum, M., Sheldon, J., & Abelson, H. (2019). From computational thinking to computational action. *Communications of the ACM, 62*(3), 34-36.

Varadarajan, A. V., Bavishi, P., Ruamviboonsuk, P., Chotcomwongse, P., Venugopalan, S., Narayanaswamy, A., ... & Webster, D. R. (2020). Predicting optical coherence tomography-derived diabetic macular edema grades from fundus photographs using deep learning. *Nature communications, 11*(1), 130.

von Wangenheim, C. G., da Cruz Alves, N., Rauber, M.F., Mayor Martins, R., & Hauck, J. C. R. (2024). Creating mobile applications with artificial intelligence adopting computational action, In Papadakis, S., Kalogiannakis, M. (eds). *Education, Development, and Intervention. Integrated Science, vol 23.* Springer, Cham., 149-165.

Waite, J., Garside, B., Whyte, R., Kirby, D., & Sentance, S. (2024, October). Experience AI: Introducing AI/ML to Grade 6–8 students in the UK. In *Proceedings of the 2024 17th International Conference on Informatics in Schools (ISSEP'24)* (pp. 28-30).

Waite, J., & Garside, B. (2023). *From ethics to engines, framing AI Literacy (including Algorithm Literacy and Data Literacy) through the SEAME framework*. Paper presented at 2024 UNESCO Global Education Meeting.

Walsh, B., Dalton, B., Forsyth, S., Haberl, E., Smilack, J., Yeh, T., ... & Sengupta-Irving, T. (2022). Aspiring for equity: Perspectives from design of AI education. In *Proceedings of the 16th International Conference of the Learning Sciences-ICLS 2022*, pp. 1771-1778. International Society of the Learning Sciences.

Wiggins, G., & McTighe, J. (2005). *Understanding by design*. Association for Supervision and Curriculum Development.